\documentclass[aps,prl,reprint,superscriptaddress]{revtex4-2}
\usepackage{graphicx}
\usepackage{color,amsmath}
\usepackage[normalem]{ulem}
\usepackage{hyperref}
\usepackage{lipsum}

\newcommand{\Vgone}{V_{\mathrm{G1}}}
\newcommand{\Vgtwo}{V_{\mathrm{G2}}}
\newcommand{\Vgthree}{V_{\mathrm{G3}}}
\newcommand{\Vs}{V_{\mathrm{S}}}
\newcommand{\Igone}{I_{\mathrm{G1}}}
\newcommand{\Igtwo}{I_{\mathrm{G2}}}
\newcommand{\Igthree}{I_{\mathrm{G3}}}
\newcommand{\Isd}{I_{\mathrm{SD}}}
\newcommand{\Ic}{I_{\mathrm{C}}}
\newcommand{\Ico}{I_{\mathrm{C0}}}
\newcommand{\Io}{I_{0}}
\newcommand{\Ir}{I_{\mathrm{R}}}
\newcommand{\Ih}{I_{\mathrm{H}}}
\newcommand{\Tq}{T_{\mathrm{Q}}}
\newcommand{\Tc}{T_{\mathrm{C}}}
\newcommand{\Rn}{R_{\mathrm{N}}}

\begin{document}
\title{On the Role of Out-of-Equilibrium Phonons in Gated Superconducting Switches}

\author{M. F. Ritter}
\affiliation{IBM Quantum, IBM Research - Zurich, S\"aumerstrasse 4, 8803 R\"uschlikon, Switzerland}

\author{N. Crescini}
\affiliation{IBM Quantum, IBM Research - Zurich, S\"aumerstrasse 4, 8803 R\"uschlikon, Switzerland}

\author{D. Z. Haxell}
\affiliation{IBM Quantum, IBM Research - Zurich, S\"aumerstrasse 4, 8803 R\"uschlikon, Switzerland}

\author{M. Hinderling}
\affiliation{IBM Quantum, IBM Research - Zurich, S\"aumerstrasse 4, 8803 R\"uschlikon, Switzerland}

\author{H. Riel}
\affiliation{IBM Quantum, IBM Research - Zurich, S\"aumerstrasse 4, 8803 R\"uschlikon, Switzerland}

\author{C. Bruder}
\affiliation{Department of Physics, University of Basel, Klingelbergstrasse 82, CH-4056 Basel, Switzerland}

\author{A. Fuhrer}
\email{afu@zurich.ibm.com}
\affiliation{IBM Quantum, IBM Research - Zurich, S\"aumerstrasse 4, 8803 R\"uschlikon, Switzerland}

\author{F. Nichele}
\email{fni@zurich.ibm.com}
\affiliation{IBM Quantum, IBM Research - Zurich, S\"aumerstrasse 4, 8803 R\"uschlikon, Switzerland}

\date{\today}

\begin{abstract}
Recent experiments suggest the possibility to tune superconductivity in metallic nanowires by application of modest gate voltages. It is largely debated whether the effect is due to an electric field at the superconductor surface or small currents of high-energy electrons. We shed light on this matter by studying the suppression of superconductivity in sample geometries where the roles of electric field and electron-current flow can be clearly separated. Our results show that suppression of superconductivity does not depend on the presence or absence of an electric field at the surface of the nanowire, but requires a current of high-energy electrons. The suppression is most efficient when electrons are injected into the nanowire, but similar results are obtained also when electrons are passed between two remote electrodes at a distance $d$ to the nanowire (with $d$ in excess of $1~\mathrm{\mu m}$). In the latter case, high-energy electrons decay into phonons which propagate through the substrate and affect superconductivity in the nanowire by generating quasiparticles. We show that this process involves a non-thermal phonon distribution, with marked differences from the loss of superconductivity due to Joule heating near the nanowire or an increase in the bath temperature. 
\end{abstract}

\maketitle

\section{Introduction}
It is commonly believed that metallic nanostructures are not affected by electric fields, as long as their size is larger than the corresponding screening length, which is typically below $5~\mathrm{nm}$. Yet, a series of experiments~\cite{DeSimoni2018,Paolucci2018,DeSimoni2019,Paolucci2019a,Puglia2020,Rocci2020} revealed a dramatic impact of gate voltages on the superconducting properties of metallic devices, such as the ambipolar quenching of the critical current. Understanding the microscopic mechanism responsible for this surprising behavior has sparked a scientific debate. First, it was suggested that an electric field can indeed penetrate a superconducting film up to the London penetration depth~\cite{DeSimoni2018}. Second, it was proposed that an electric field might perturb the polarization of atomic orbitals at the metal surface, and this would affect the superconducting properties in the bulk~\cite{Mercaldo2020,Paolucci2019b}. Third, studies of the switching probability distribution in metallic nanowires suggested an interplay between an electric field and superconducting phase slips~\cite{Puglia2020}.

\begin{figure*}
 \includegraphics[width=2\columnwidth]{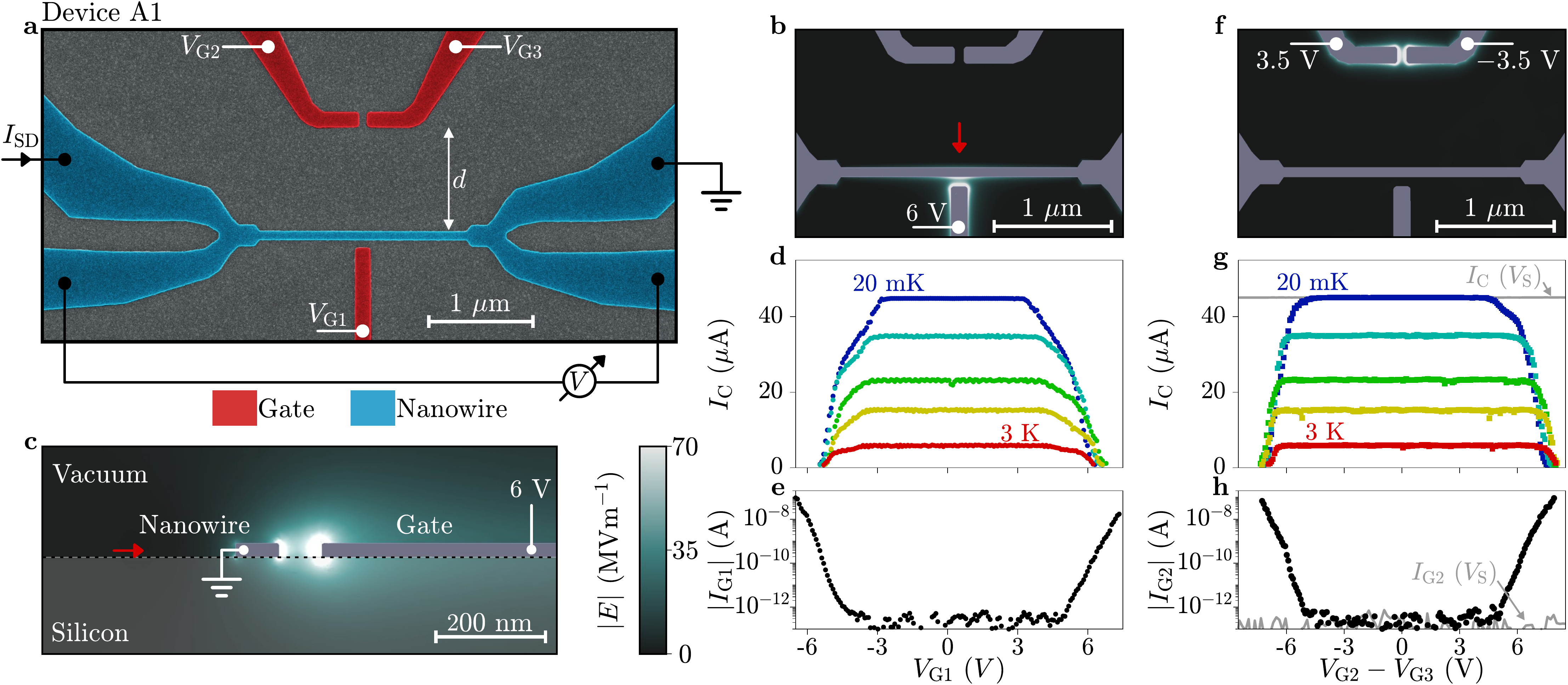}
 \caption{(a) False-color scanning electron micrograph of Device~A1, wbith simplified measurement configuration. The nanowire under investigation is depicted blue and the gates red. (b) Finite element simulation of the electric field magnitude $|E|$ for $\Vgone=6~\mathrm{V}$. We show a slice of the three dimensional simulation on a plane elevated $10~\mathrm{nm}$ from the Si substrate. (c) Same as in (b), but for a plane perpendicular to the substrate and intersecting gate~1. The red arrow indicates the direction of the cut in (b). (d) Critical current $\Ic$ in Device~A1 as a function of gate voltage $\Vgone$ for temperatures $T$ of $20~\mathrm{mK}$ (blue), $1.5~\mathrm{K}$, $2.1~\mathrm{K}$, $2.5~\mathrm{K}$ and $3~\mathrm{K}$ (red). (e) Gate current $\Igone$ as a function of $\Vgone$ measured at $T=20~\mathrm{mK}$ simultaneously to the data in (d). (f) finite element simulation as in (b), but calculated for $\Vgtwo-\Vgthree=7~\mathrm{V}$. (g) Critical current $\Ic$ in Device~A1 as a function of gate voltage difference $\Vgtwo-\Vgthree$ for temperatures as in (d) (markers), together with $\Ic$ as a function of $\Vs=2\Vgtwo=2\Vgthree$, representing twice the voltage applied to both gates simultaneously (gray line). (h) Current $\Igtwo$ flowing from gate 2 as a function of voltage difference $\Vgtwo-\Vgthree$. In this configuration $\Igtwo=-\Igthree$ within experimental error. Gate current $\Igtwo$ as a function of $\Vs$ is shown in gray.}
 \label{fig:1}
\end{figure*}

In Ref.~\onlinecite{Ritter2021}, we have reproduced the most distinctive features of previous experiments using TiN, Nb and Ti nanowires. In our samples, the critical current suppression was always accompanied by a current flowing between gate and nanowire. In these experiments the gate current is carried by electrons with energies of several eV, which is orders of magnitude larger than the superconducting energy gap in the nanowires. We concluded that the emission of relatively few electrons lead to an avalanche of quasiparticles, which effectively quench the critical current~\cite{Engel2013}. This hypothesis was supported by tunneling spectroscopy experiments~\cite{Alegria2021}, which highlighted a non-thermal increase in quasiparticle population as a gate voltage was applied. Further work also demonstrated a correlation between onset of gate currents and suppression of superconducting properties~\cite{Golokolenov2021,Catto2021}. However, open questions still remain. For example, in a scenario where injection of high-energy electrons controls the critical current suppression, a marked asymmetry would naively be expected between injecting high-energy electrons into the nanowire (negative gate voltage) and extracting electrons from the nanowire at the Fermi energy (positive gate voltage), and having them relax either in the substrate or in the gate electrode. In view of possible technological applications of this phenomenon, such as the realization of voltage-controlled superconducting switches and resonators, unraveling the microscopic mechanisms behind these observations is a pressing task.

In this Article, we present conclusive evidence linking the quench of superconductivity in our nanowires to the relaxation of high-energy electrons, and not to the presence of electric fields at the superconductor surface. In particular, we investigate the effect of high-energy electrons flowing into the nanowire, out from the nanowire, or between two remote gate electrodes in the vicinity of the nanowire. Detailed measurements reveal that superconductivity is most efficiently suppressed when a current is injected into the nanowire. However, a qualitatively similar critical current suppression is observed when high-energy electrons flow near the nanowire, without any current or electric field directly reaching the nanowire itself. The non-local nature of the observed effect is consistent with energy relaxation of electrons by phonon emission in the substrate. Due to their relatively high energy, phonons generate quasiparticles in the superconductors and efficiently quench the critical current in our devices. At cryogenic temperatures phonons can propagate over considerable distances in the crystalline silicon substrate before thermalizing. The effect that we describe is therefore markedly different from the situation where a local temperature increase is produced by a resistive heater. Our observations question existing interpretations and theories based on electric fields and contribute towards understanding the complex interactions between out-of-equilibrium phenomena in solids and performance of superconducting hardware.

\section{Results and Discussion}
Seven TiN nanowires on Si substrates were investigated during this work. All nanowires had a length of $2~\mathrm{\mu m}$, a width of $80~\mathrm{nm}$ and a height of $20~\mathrm{nm}$. At low temperature, devices showed critical currents $\Ic$ between $42$ and $45~\mathrm{\mu A}$, a retrapping current $\Ir=1.0~\mathrm{\mu A}$ and a normal state resistance $\Rn\sim1750~\mathrm{\Omega}$, consistent with previous work~\cite{Ritter2021}. The consistency of these values demonstrates that the nanowires were homogeneous and not characterized by accidental weak links. Further details on sample fabrication and basic characterization are reported in Ref.~\onlinecite{Ritter2021} and in the Methods section. Here, we present results from four devices, referred to as Device A1, A2, B and C, respectively. Three additional devices, used as references, are shown in more detail in the Supplementary Information~\cite{Supplement}.

Figure~\ref{fig:1}(a) shows a false-color scanning electron micrograph of Device~A1, together with a schematics of the measurement configuration. Device~A1 consists of a nanowire (blue) and three gates (red). Gate~1, controlled by the voltage $\Vgone$, was separated from the nanowire by a gap of $80~\mathrm{nm}$. Gates~2 and~3, controlled by voltages $\Vgtwo$ and $\Vgthree$ respectively, were separated from each other by $80~\mathrm{nm}$ and from the nanowire by a distance $d=1~\mathrm{\mu m}$. A similar device, named Device~A2, had $d=80~\mathrm{nm}$ and is presented in the Supplementary Information~\cite{Supplement}.

We first discuss the response of Device~A1 to a side gate voltage $\Vgone$, similar to previous work~\cite{DeSimoni2018,Alegria2021,Ritter2021}. The electric field distribution in this configuration was calculated using three-dimensional finite element simulations (see Methods Section). Figure~\ref{fig:1}(b) shows the field magnitude $|E|$ on a plane $10~\mathrm{nm}$ above the substrate for $\Vgone=6~\mathrm{V}$. Figure~\ref{fig:1}(c) represents $|E|$ on a plane perpendicular to both the substrate and the wire axis, and intersecting the gate (see red arrow in Fig.~\ref{fig:1}(b)). To better highlight the field distribution, the color scale was saturated to $|E|=70~\mathrm{MVm^{-1}}$. The highest $|E|$ in our simulations was below $|E|=500~\mathrm{MVm^{-1}}$, which is several orders of magnitude smaller than typical electric fields required to perturb superconductivity in a metallic device~\cite{Glover1960,Choi2014,Piatti2017}. Figure~\ref{fig:1}(d) shows the experimentally measured $\Ic$ as a function of $\Vgone$, for temperatures ranging from $20~\mathrm{mK}$ (blue) to $3~\mathrm{K}$ (red). Figure~\ref{fig:1}(e) shows the gate current $\Igone$ measured simultaneously to the data in Fig.~\ref{fig:1}(d). Consistent with previous observations~\cite{Ritter2021,Golokolenov2021}, the decrease of $\Ic$ was correlated to the onset of $\Igone$, and the initial decrease in $\Ic$ took place for $\Igone<1~\mathrm{pA}$. Furthermore, $|\Igone|$ was found to increase exponentially with $\Vgone$ and to be approximately symmetric around $\Vgone=0$.

We now discuss the dependence of $\Ic$ on a differentially applied voltage $\Vgtwo-\Vgthree$, with $\Vgtwo=-\Vgthree$. Figure~\ref{fig:1}(f) shows the numerically computed electric field for $\Vgtwo-\Vgthree=7~\mathrm{V}$. As expected, $|E|$ is strongly confined between Gates~2 and 3. If superconductivity in the nanowire were controlled by the electric fields, this configuration should result in negligible effects on $\Ic$. Strikingly, quenching of the supercurrent occurred also in this situation, as shown in Fig.~\ref{fig:1}(g). Figure~\ref{fig:1}(h) shows the current $\Igtwo$ flowing from Gate~2 (we found $\Igtwo=-\Igthree$ within experimental error). Remarkably, the suppression of $\Ic$ was strongly correlated to the onset of $\Igtwo$, despite no measurable gate current reached the nanowire and electric fields between gate and nanowire were negligible.

To test whether residual electric fields were relevant, we also measured $\Ic$ with Gate~2 and~3 biased at the same voltage ($\Vgtwo=\Vgthree$). In Fig.~\ref{fig:1}(g) we plot $\Ic$ as a function of the quantity $\Vs = 2\Vgtwo = 2\Vgthree$ (solid gray line in Fig.~\ref{fig:1}(g)) as, at any one point in this plot, the absolute voltages $|\Vgtwo|$ and $|\Vgthree|$ on the gate electrodes are identical, and we estimate $|E(\Vgtwo=\Vgthree)|\gtrsim |E(\Vgtwo=-\Vgthree)|$ at the nanowire surface. Nevertheless, no current was detected between gates and nanowire for symmetrically applied gate voltages (see gray curve in Fig.~\ref{fig:1}(h)) and $\Ic$ was not perturbed. These results further corroborate our findings that high-energy electrons, and not electric fields, are responsible for the suppression of $\Ic$. Similar results obtained with Device~A2 are presented in the Supplementary Information~\cite{Supplement}.

\begin{figure*}
 \includegraphics[width=2\columnwidth]{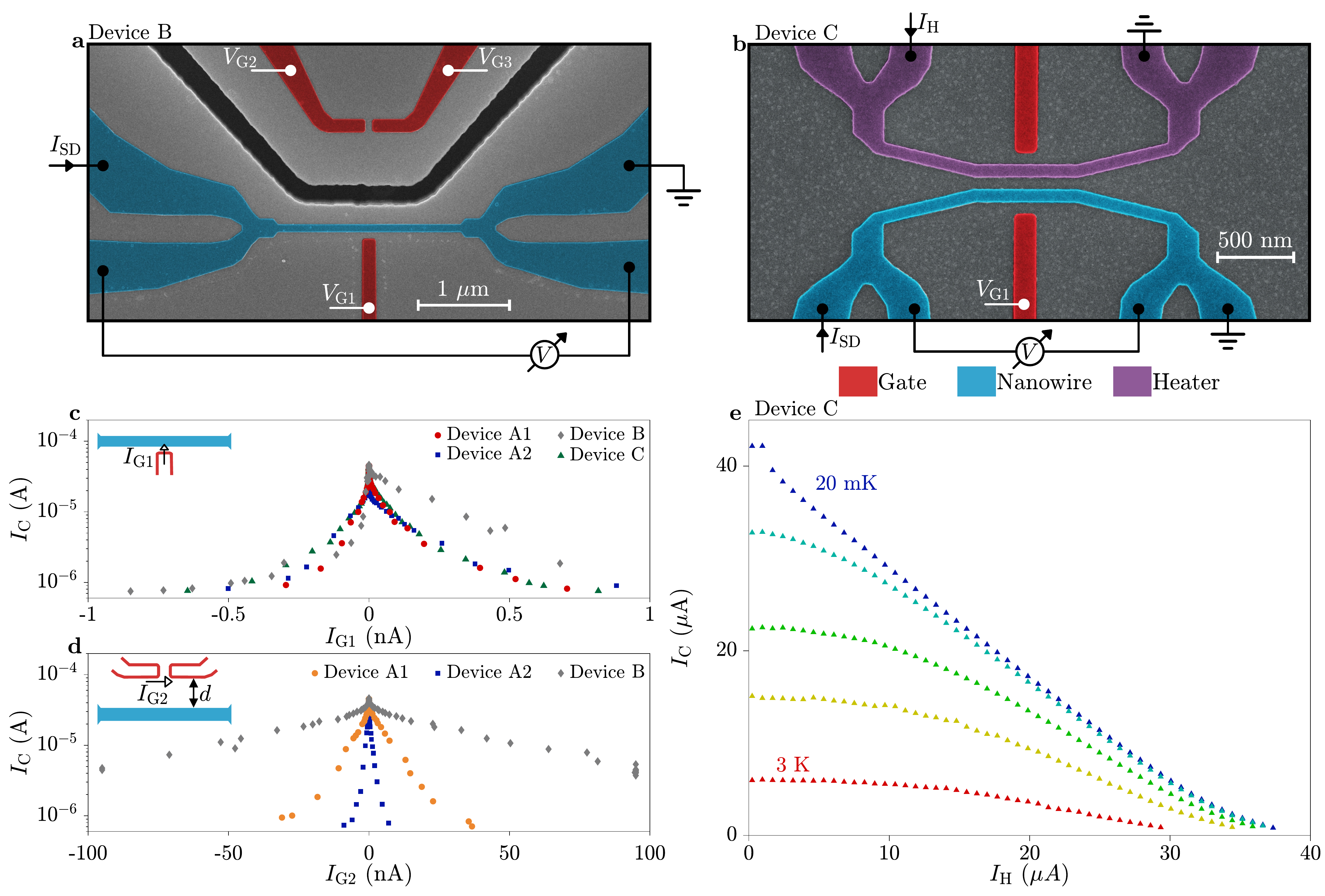}
 \caption{(a) False-color scanning electron micrograph of Device~B. The nanowire is depicted blue, the gates red and the trench appears black. The trench is $510~\mathrm{nm}$ deep, $200~\mathrm{nm}$ wide and has a total length of $80~\mathrm{\mu m}$. (b) False-color scanning electron micrograph of Device~C. The nanowire under investigation is depicted blue, the gates red and the heater nanowire purple. (c) Plot of $\Ic$ as a function of $\Igone$ for all the devices presented in the Main Text. (d) Plot of $\Ic$ as a function of the remote gate current $\Igtwo$ for Devices~A1 ($d=1~\mathrm{\mu m}$), A2 ($d=80~\mathrm{nm}$) and~B ($d=1~\mathrm{\mu m}$ plus a trench). (e) Critical current in Device~C as a function of heater current $\Ih$ for temperatures as in Fig.~\ref{fig:1}(d).}
 \label{fig:2}
\end{figure*}

Overall, experiments and numerical simulation presented in Fig.~\ref{fig:1} demonstrate that the suppression of superconductivity takes place irrespective of electric fields at the nanowire surface. Instead it requires the flow of high-energy electrons in the surroundings of the device. This is the first conclusion of our work. Furthermore, the remote action of $\Vgtwo-\Vgthree$ on $\Ic$ points to the existence of an efficient energy transfer mechanism triggered by the flow of $\Igtwo$. We now analyze the origin of this remote action more carefully using Devices~B and~C, shown in Fig.~\ref{fig:2}(a) and (b), respectively. Device~B is identical to Device~A1, except for the presence of a $510~\mathrm{nm}$ deep, $200~\mathrm{nm}$ wide and $80~\mathrm{\mu m}$ long trench etched into the substrate between the remote gates and the nanowire. Device~C consists of two parallel TiN nanowires separated by a distance of $80~\mathrm{nm}$. Each nanowire was controlled by a nearby gate (red). We measured the critical current of one of the two nanowires (blue) while the second one (purple) was set in the resistive state and was traversed by a DC current $\Ih$, resulting in Joule heating.

Figure~\ref{fig:2}(c) and (d) summarize the behavior of our devices in terms of $\Ic$ as a function of $\Igone$ and $\Igtwo$, respectively. The full dataset is presented in the Supplementary Information~\cite{Supplement}. The dependence on $\Igone$ (see Fig.~\ref{fig:2}(c)) is similar in all devices, with a faster suppression of $\Ic$ for $\Igone<0$. Due to the exponential dependence of $\Igone$ on $\Vgone$, this asymmetry is hard to spot in Figs.~\ref{fig:1}(d) and (e). We further notice that Device~B (gray diamonds) exhibited a particularly slow decay of $\Ic$ for $\Igone>0$. We will discuss possible causes for this asymmetry below.
Figure~\ref{fig:2}(d) reveals that $\Igtwo$ is significantly less effective in suppressing $\Ic$ than $\Igone$. Furthermore, Device~A2 (blue squares, $d=80~\mathrm{nm}$), was more efficient than Device~A1 (orange circles, $d=1~\mathrm{\mu m}$), which was more efficient than Device~B (gray diamonds, $d=1~\mathrm{\mu m}$ plus an etched trench). In the case of Device~B, the maximum $\Igtwo$ allowed in our setup ($100~\mathrm{nA}$) was not sufficient to reach $\Ic=0$. Altogether, these results demonstrate that most of the remote action of $\Igtwo$ on $\Ic$ is mediated by the substrate, i.e. the high-energy electrons relax by emitting phonons, which travel through the substrate and affect superconductivity in the nanowire. This is the second main conclusion of our work.

We now discuss the properties of the generated phonons in more detail. In particular, we compare their effect on $\Ic$ to that of heat generated by a resistive conductor placed $80~\mathrm{nm}$ from the superconducting nanowire. These experiments were performed with Device~C, shown in Fig.~\ref{fig:2}(b). The dependence of $\Ic$ on the heater current $\Ih$ is shown in Fig.~\ref{fig:2}(e) for various temperatures. As expected, Joule heating eventually resulted in the suppression of $\Ic$. However, the current required to reach $\Ic=0$ was several orders of magnitude higher than in the configurations where a gate voltage was applied.

\begin{figure}
 \includegraphics[width=\columnwidth]{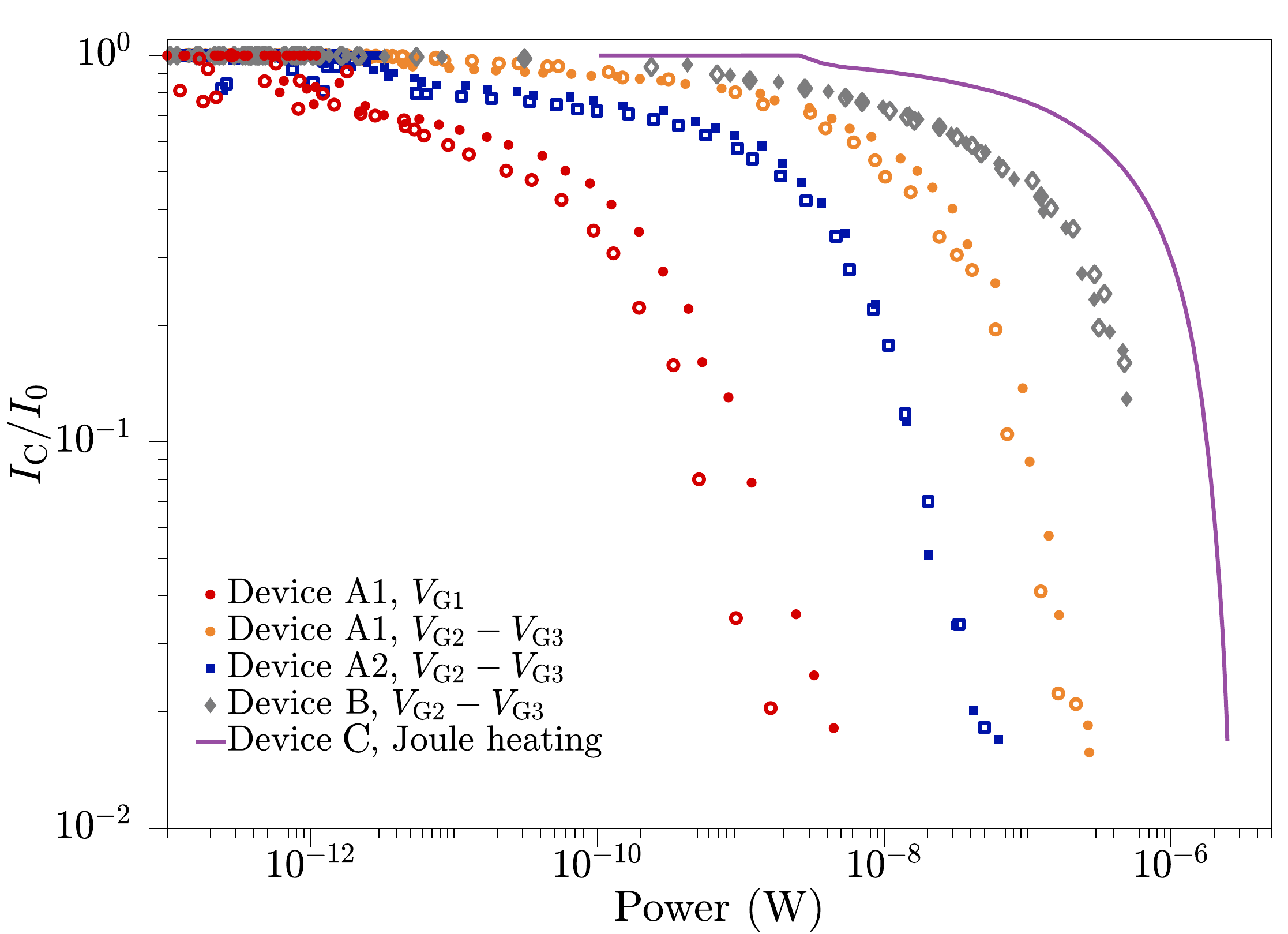}
 \caption{Normalized critical current $\Ic/\Io$ as a function of input power for various devices and experimental configurations (see legend). Full and empty markers define positive and negative gate polarity, respectively. The solid purple line indicates the dependence as a function of Joule heating in Device~C.}
 \label{fig:3}
\end{figure}
Figure~\ref{fig:3} provides a comparison between the devices presented above in terms of the suppression of normalized critical currents $\Ic$ as a function of dissipated power. For each measurement configuration, we distinguish the case of positive and negative voltage bias with full and empty markers, respectively. $\Ic$ is more efficiently suppressed when a voltage bias $\Vgone$ is applied to a gate directly facing the nanowire (red dots). In this case, the dissipated power is calculated as $\Igone\Vgone$. When a remote current $\Igtwo$ flows, the power is calculated as $\Igtwo(\Vgtwo-\Vgthree)$. Suppressing $\Ic$ by means of Joule heating with a resistive conductor (purple line) required a significantly higher power $\Ih^2\Rn$ than the other configurations. As noted above, the dependence on $\Igone$ (red circles) shows a difference between positive and negative gate polarity, with the negative polarity being four times more power efficient in suppressing $\Ic$ with respect to the positive one.

\begin{figure}
 \includegraphics[width=\columnwidth]{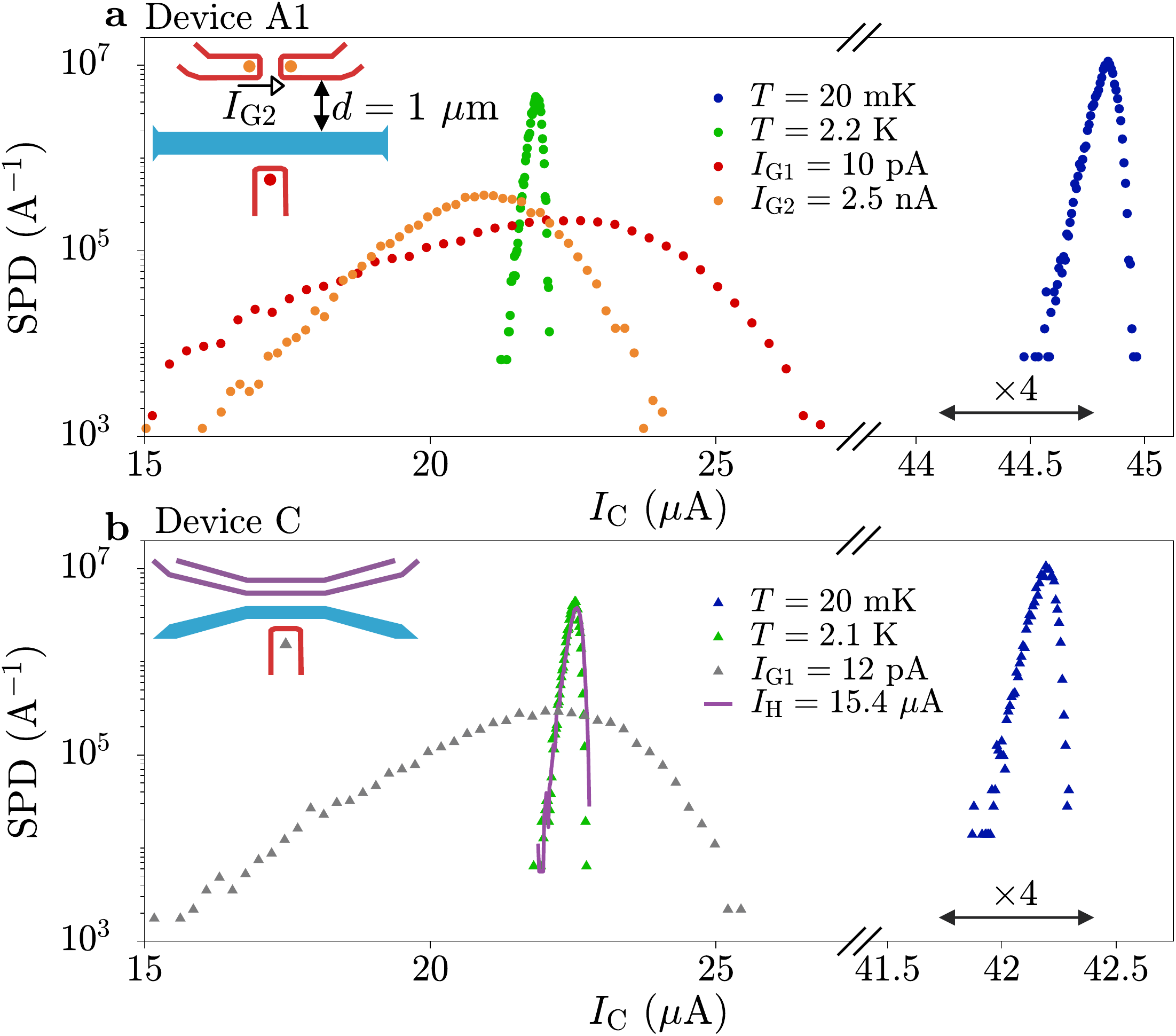}
 \caption{(a) Switching probability distribution (SPD) in Device~A1 as a function of source-drain current $\Isd$. Blue and green circles are obtained at zero gate voltage and for $T=20~\mathrm{mK}$ and $T=2.2~\mathrm{K}$, respectively. Red circles are obtained for $\Igone=10~\mathrm{pA}$, ($\Vgone=5.85~\mathrm{V}$), orange dots for $\Igtwo=2.5~\mathrm{nA}$ ($\Vgtwo-\Vgthree=7.25~\mathrm{V}$). Note that the horizontal axis is interrupted and the high current region is horizontally expanded by a factor 4. (b) As in (a), but for Device~C. Blue and green circles are obtained at zero gate voltage and for $T=20~\mathrm{mK}$ and $T=2.1~\mathrm{K}$, respectively. Gray markers are obtained for $\Igone=12~\mathrm{pA}$ ($\Vgone=5.2~\mathrm{V}$), the purple line is obtained for a heater current $\Ih=15.4~\mathrm{\mu A}$.}
 \label{fig:4}
\end{figure}

We have shown that Joule heating is orders of magnitude less efficient in suppressing $\Ic$ of our nanowires than a current of high-energy electrons. In addition to these quantitative differences, we also identified striking qualitative differences in the switching probability distributions (SPDs) of our devices. The SPD is the probability for a switch from superconducting to resistive state to occur per unit of source-drain current. The SPD has proven to be a powerful tool to study Josephson junctions and metallic nanowire properties that are hard to access with standard transport measurements~\cite{Bezryadin2012a,Puglia2020}. Figures~\ref{fig:4}(a) and (b) show the SPDs of Devices~A1 and~C, respectively, under various experimental conditions. For these experiments, the source-drain current was swept $20,000$ times from zero to $49~\mathrm{\mu A}$. For each sweep, the source-drain current value at which a switch to the resistive state occurred was recorded. At low temperature and zero gate voltage, Device~A1 exhibited a sharp SPD (blue markers), with a standard deviation of $47~\mathrm{nA}$. At a temperature of $2.2~\mathrm{K}$ (green markers) the SPDs had their maximum at half of its low temperature value, with a standard deviation of $100~\mathrm{nA}$. More detailed analysis, reported in the Supplementary Information~\cite{Supplement}, revealed that the switching mechanisms at $20~\mathrm{mK}$ and $2.2~\mathrm{K}$ are consistent with quantum phase slips and thermal fluctuations, respectively. Much broader SPDs were obtained by applying a gate leakage current $\Igone=10~\mathrm{pA}$ (red markers). The finding that the application of a gate voltage results in much broader SPDs than increasing the bath temperature (for equal $\Ic$ value) is consistent with the observations in Ref.~\onlinecite{Puglia2020}. However, we show that a similarly broad SPD is also obtained by applying a remote current $\Igtwo=2.5~\mathrm{nA}$ (orange circles in Fig.~\ref{fig:4}(a)), that is without any electric field or current reaching the nanowire. Using Device~C (Fig.~\ref{fig:4}(b)), we compare the SPD obtained when $\Ic$ is suppressed by $50\%$ either by Joule heating (solid purple line) or by increasing the bath temperature to $2.1~\mathrm{K}$ (green triangles). The two results are indistinguishable, indicating that a resistive heater indeed affects superconductivity in the same way as an increase in bath temperature, but in a totally different manner than a current of high-energy electrons (gray triangles). The difference between SPDs obtained at high temperature (green markers) and finite gate voltage (red markers) led the authors of Ref.~\onlinecite{Puglia2020} to exclude the presence of electrical currents. This conclusion was however reached under the assumption that a gate current causes heating similar to an increase of the bath temperature. Our results demonstrate instead that a current of high-energy electrons perturbs the superconducting properties of nanowires in a way that is qualitatively and quantitatively distinct from a bare temperature increase, even if the current does not flow into the nanowire but only in its surroundings. This is the third main conclusion of our work.

Our observations are consistent with the phenomenology of phonon generation by hot electrons in the substrate. First, we note that phonons with energies above the superconducting gap ($500~\mathrm{\mu eV}$ for TiN~\cite{Pracht2012}) are well known to affect superconducting devices~\cite{Eisenmenger1967,Dayem1971,Singer1976,Ioffe2004}. Second, electrons accelerated by high electric fields in Si undergo a series of relaxation events over time scales below $1~\mathrm{ns}$ and on mean free paths below $10~\mathrm{nm}$. Such relaxation most likely happens by emission of  optical and acoustic phonons~\cite{Brunetti1981,Fischetti1995,Sadasivam2017,Tanimura2019}. Phonons in Si have a maximum energy of the order of $50~\mathrm{meV}$, which means that a single electron with energies of a few~eV can generate a large amount of phonons~\cite{Sinha2005,Pop2006} as it travels between two metallic electrodes. At temperatures below 3~K phonons in Si have long mean free paths (up to $1~\mathrm{\mu m}$ ~\cite{Ju1999,Anufriev2020}) and even longer thermalization lengths. It is therefore expected that the emitted phonons reach the nanowire in an out-of-equilibrium state. The electronic mean free path in Si decreases as $|E|$ increases~\cite{Brunetti1981,Fischetti1995}, resulting in intense phonon emission close to the metal electrodes, independent of gate voltage polarity (see simulation in Fig.~\ref{fig:1}(c))~\cite{Sinha2005}. This may be the reason for the more efficient suppression of $\Ic$ when a current is either injected or extracted from the nanowire (see Fig.~\ref{fig:2}(c)) compared to the case where a current flows between two gates near the nanowire (see Device~A2 in Fig.~\ref{fig:2}(d)). The faster suppression of $\Ic$ observed for $\Vgone<0$ (Fig.~\ref{fig:2}(c)) is due to high-energy electrons reaching the nanowire and directly creating quasiparticles in the superconductor through electron-electron interaction~\cite{Roukes1985,Wellstood1994,Engel2013}. Future work might use more complex geometries to map out angular anisotropies in the phonon emission and absorption processes.

In Device~B we noticed an anomalously large asymmetry in the parametric plot of $\Ic$ vs. $\Igone$ (see Fig.~\ref{fig:2}(c)). We have confirmed with three reference devices (see Supplementary Information~\cite{Supplement}) that such an asymmetry is a robust feature which arises following the fabrication steps required to etch trenches into the substrate (see Methods Section). Similarly, the efficiency of the remote action of $\Igtwo$ slightly decreased after additional fabrication, even when trenches were not etched (see Supplementary Information~\cite{Supplement}). Interestingly, no other sample parameters were affected by the additional fabrication steps. These results suggest that some of the out-of-equilibrium processes taking place in our device are sensitive to the surface treatment of the samples. Measuring Device~B, we have shown that out-of-equilibrium phonons are the main responsible for the remote action of $\Igtwo$ on $\Ic$. However, our work does not exclude the presence of additional energy relaxation mechanisms which contribute, together with phonons, to the suppression of $\Ic$, such as photon emission. Previous works detected photons in a variety of devices as a result of tunneling events~\cite{Lambe1976,Gimzewski1988,Uehara1992,Parzefall2015,Doderer2019} as well as \textit{bremsstrahlung} and carrier recombination of high-energy electrons~\cite{Bude1992,Lacaita1993}. It is also well known that superconducting nanowires~\cite{Natarajan2012} and Josephson junctions~\cite{Walsh2021} are highly sensitive to the impact of high-energy photons. Both phonon and photon transport may be affected by the additional fabrication steps for the trenching e.g. by a change in surface roughness or dielectric properties. Further work may be needed to quantitatively address the contributions of phonons and photons in the remote response of superconducting nanowires as ours. 

\section{Conclusions}
In conclusion, we performed a comprehensive study of the mechanism responsible for the suppression of critical currents in metallic nanowires in the presence of large gate voltages. We have shown that all previously reported features which were attributed to the electric field on the superconductor can be obtained in the absence of electric fields. Our data indicates that critical currents are suppressed as a consequence of the relaxation of high-energy electrons, either in the substrate or in the electrodes. Our results elucidate the mechanism behind the ambipolar suppression of $\Ic$ as a function of gate voltage (see Fig.~\ref{fig:1}(d)), which was not fully explained in previous works~\cite{Ritter2021,Alegria2021,Golokolenov2021}. The ambipolar suppression of $\Ic$ requires both an approximately symmetric gate current, which is experimentally observed (see Fig.~\ref{fig:1}(e)), and an efficient energy equilibration mechanism between gate and nanowire. Energy equilibration is dominated, in our devices, by energetic phonons spreading through the substrate over distances in excess of $1~\mathrm{\mu m}$. While this remote action may pose a limit to device integration density, it could also open new paths for device design. For example, it could allow for efficient superconducting switches~\cite{Morpurgo1998,McCaughan2016,McCaughan2019,Wagner2020} which do not require injection of electrons into the switching element, but are instead mediated by high-energy phonons that are guided towards a switching element. Also, it opens new possibilities to investigate the interplay between out-of-equilibrium phenomena, resulting quasiparticle generation, and superconducting quantum hardware.

\textbf{Acknowledgments}
We are grateful to A.~Pushp, B.~Madon and M.A.~Mueed for deposition of the TiN films and to V. Geshkenbein for fruitful discussions. We thank the Cleanroom Operations Team of the Binnig and Rohrer Nanotechnology Center (BRNC) for their help and support. F.~Nichele acknowledges support from the European Research Commission, grant number 804273.  A.~Fuhrer acknowledges support from the Swiss National Science Foundation through Grant No. 200021\_188752.

\section{Methods}
\textbf{Sample Fabrication.}
A $20~\mathrm{nm}$ thick TiN film was sputtered on a Si substrate. The Si substrate used for this work was intrinsic and became insulating at temperatures below $100~\mathrm{K}$. Prior to TiN deposition, the Si chip was immersed in a buffered HF solution for removal of native oxides. The TiN film showed a critical temperature of $3.7~\mathrm{K}$ and a resistivity of $68~\mathrm{\Omega}$ per square. Devices were defined by electron beam lithography on a negative HSQ (hydrogen silsesquioxane) resist and dry etching in a HBr plasma. The HSQ resist was then removed by immersion in HF. Devices were contacted by Ti/Au bond pads defined by optical lithography and metal evaporation.
Some devices were further processed after deposition of the bond pads. In this case a $2~\mathrm{nm}$ $\mathrm{Si_3N_4}$ and $210~\mathrm{nm}$ $\mathrm{SiO_2}$ hard mask were deposited by atomic layer deposition and plasma enhanced chemical vapor deposition, respectively. A trench was defined in the hard mask with electron beam lithography and a CSAR resist, standard development and reactive ion etching of the $\mathrm{SiO_2}$ layer. The Si substrate was further etched in an inductively coupled HBr plasma. Finally, the hard mask was etched in buffered HF.

\textbf{Electrical Measurements.}
Measurements were performed in a dilution refrigerator with a base temperature of $20~\mathrm{mK}$. Critical currents $\Ic$ were measured by applying a sawtooth wave $\Isd$ signal with amplitude $49~\mathrm{\mu A}$ and repetition rates between $33$ and $133~\mathrm{Hz}$ while voltages $V$ across the nanowires were recorded by a digital oscilloscope. The measurement setup was synchronized so that a switch from zero to finite voltage in the oscilloscope could be related to the source-drain current at which the switch occurred. This technique allowed us to reliably extract critical currents down to $700~\mathrm{nA}$. Critical currents presented in Fig.~\ref{fig:1} were obtained by averaging $108$ such switching events. Sporadic fluctuations of $\Ic$ visible at $T\geq1.5~\mathrm{K}$ are associated to instabilities of the temperature controller. Switching probability distributions presented in Fig.~\ref{fig:4} were obtained by recording $20,000$ switches over a time interval of 10~minutes. In order to keep the nanowire potential constant while $\Isd$ varies, $\Isd$ was generated by sourcing two synchronized sawtooth waves with opposite polarity into $163~\mathrm{k\Omega}$ resistors placed at both ends of the nanowire (which add to the existing $2.2~\mathrm{k\Omega}$ line resistance). Gate voltages were applied via high-precision source-measure units, which recorded the current flowing into the gate contacts. Gate current data as in Fig.~\ref{fig:1}(e) and (h) were obtained after subtracting linear components ranging between $1$ and $5~\mathrm{pA~V^{-1}}$, as discussed in Ref.~\onlinecite{Ritter2021}. Such resistive contributions are attributed to spurious leakage paths in our setup.

\textbf{Electrostatic Simulations.}
Electric field distributions presented in Fig.~\ref{fig:1} were produced with finite element 3D electrostatic simulations performed with ANSYS Maxwell. A substrate permittivity of 12 was assumed in order to resemble the electromagnetic properties of silicon, and its thickness was set to $1~\mathrm{\mu m}$. The metallic layer comprising the nanowire and the gate electrodes was modeled as a $20~\mathrm{nm}$ thick perfect conductor. The upper edges of the structures were filleted with a radius of $3~\mathrm{nm}$. The geometry of nanowire and gates was generated from the same layout file used for the electron beam lithography of the devices. The fields shown in Figs.~\ref{fig:1}(b) and (f) are slices of the three-dimensional simulation taken at half the height of the nanowire. Figure~\ref{fig:1}(c) is taken perpendicular to the substrate and intersecting the gate electrode. The color scale was saturated to a maximum value of $70~\mathrm{MVm^{-1}}$ to evidence the field distribution, while the full scale reached up to $500~\mathrm{MVm^{-1}}$.

\section{Data Availability}
The data that support the findings of this study are available upon reasonable request from the corresponding author.

\bibliography{Bibliography}

\clearpage
\newpage

\newcounter{myc} 
\renewcommand{\thefigure}{S.\arabic{myc}}

\title{Supplementing Information}
\maketitle
\section{Supplementing Information~1: Device~A2}
\begin{figure}
 \includegraphics[width=\columnwidth]{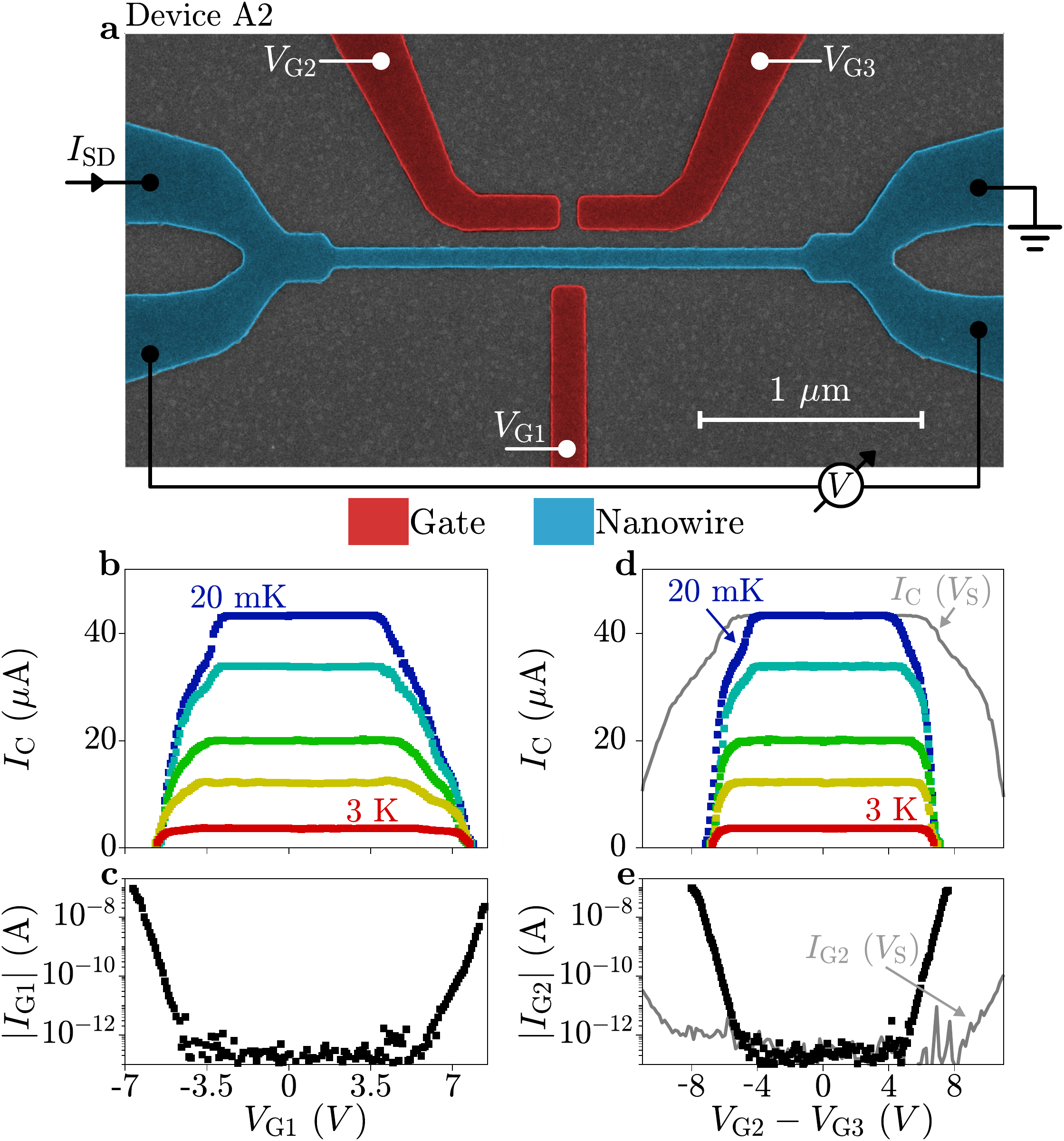}
 \caption{(a) False-color scanning electron micrograph of Device~A2 and simplified measurement configuration. The nanowire under investigation is depicted blue and the gates red. (b) Critical current $\Ic$ as a function of gate voltage $\Vgone$ at temperatures $T$ of $20~\mathrm{mK}$ (blue), $1.5~\mathrm{K}$, $2.1~\mathrm{K}$, $2.5~\mathrm{K}$ and $3.0~\mathrm{K}$ (red). (c) Gate current $\Igone$ as a function of $\Vgone$ at $T=20~\mathrm{mK}$. (d) Gate current $\Igtwo$ as a function of $\Vgtwo-\Vgthree$ for the same temperature values as in (b) (markers) together with $\Ic$ as a function of the parameter $\Vs=2\Vgtwo=2\Vgthree$. (e) Gate current $\Igtwo$ as a function of gate voltage difference $\Vgtwo-\Vgthree$ measured at $T=20~\mathrm{mK}$ (black markers), together with the current $\Igtwo$ as a function of $\Vs$.}
 \label{fig:S1}
\end{figure}

Device~A2 was lithographically equivalent to Device~A1, shown in Fig.~1(a) of the Main Text, except for the distance $d$ between gates~2,3 and the nanowire ($d=1~\mathrm{\mu m}$ in Device~A1 and $d=80~\mathrm{nm}$ in Device~A2). A false-color image of Device~A2 is shown in Fig.~\ref{fig:S1}, together with a simplified measurement schematic. The nanowire under study is depicted blue, the three gates red.
Critical current $\Ic$ and gate current $\Igone$ as a function of $\Vgone$ are shown in Fig.~\ref{fig:S1}(b) and (c) respectively. Their characteristics are almost identical to those of Device~A1 (see Fig.~1(d,e) of the Main Text). The dependence on $\Vgtwo-\Vgthree$ is depicted in Figs.~\ref{fig:S1}(d) and (e). Suppression of $\Ic$ was again correlated with the increase of the current $\Igtwo$. This time, sweeping gates~2 and 3 at the same voltage resulted in a partial suppression of $\Ic$ (gray line in Fig.~\ref{fig:S1}(d) shows $\Ic$ as a function of the parameter $\Vs=2\Vgtwo=2\Vgthree$) correlated with currents flowing from gates 2 and 3 into the nanowire (gray line in Fig.~\ref{fig:S1}(e)). The fact that $\Ic$ was affected at higher voltages for equal gate biases $\Vgtwo=\Vgthree$ with respect to the asymmetric bias configuration ($\Vgtwo=-\Vgthree$) (markers) speaks against any effect linked to electric fields between gates and nanowire.

\setcounter{myc}{1}

\section{Supplementing Information~2: Devices B and C}
\begin{figure}
 \includegraphics[width=\columnwidth]{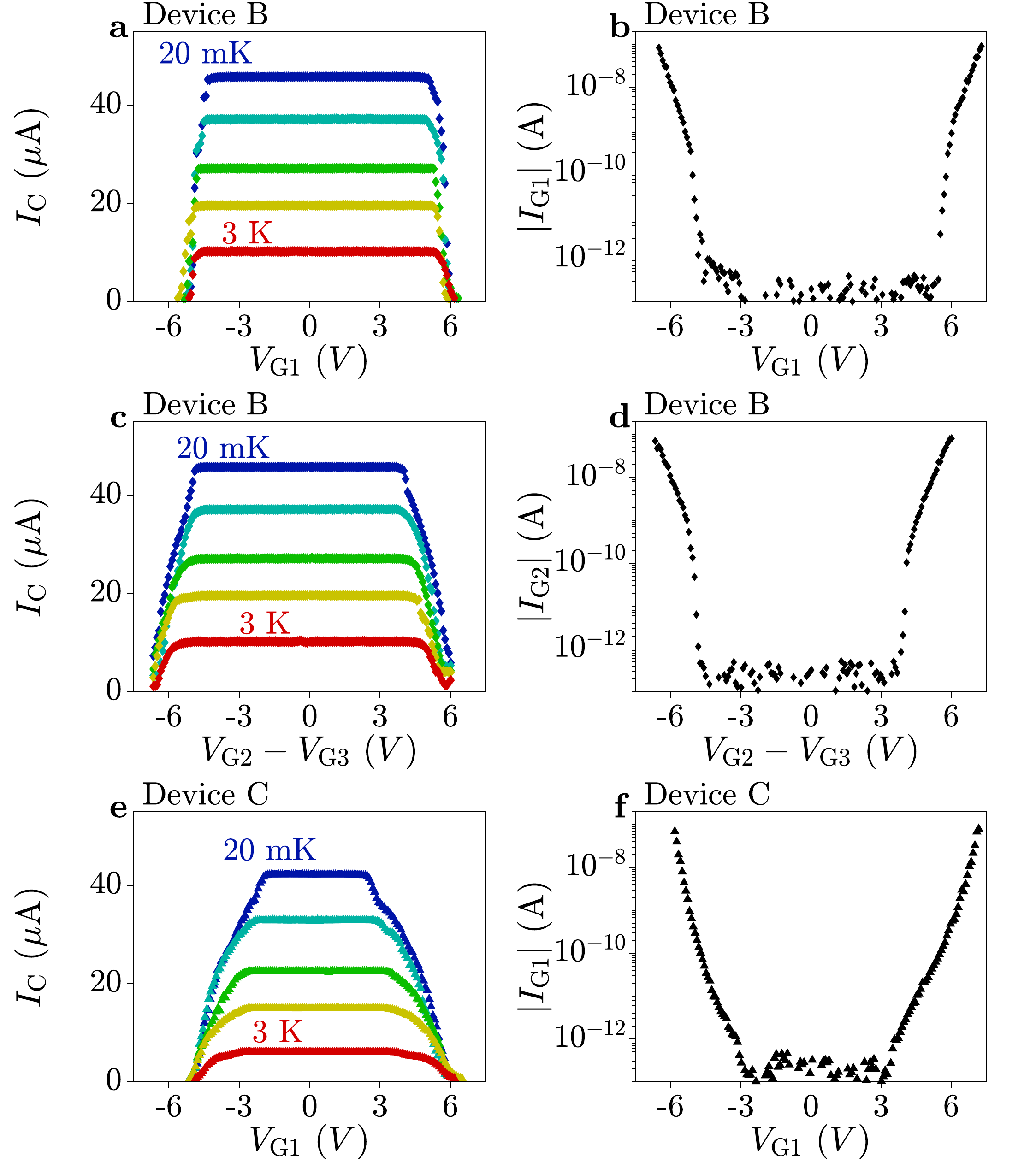}
 \caption{(a) Critical current $\Ic$ of Device~B as a function of gate voltage $\Vgone$ at temperatures $T$ of $20~\mathrm{mK}$ (blue), $1.5~\mathrm{K}$, $2.1~\mathrm{K}$, $2.5~\mathrm{K}$ and $3.0~\mathrm{K}$ (red). (b) Gate current $\Igone$ of Device~B as a function of $\Vgone$ at $T=20~\mathrm{mK}$. (c) Critical current $\Ic$ as a function of $\Vgtwo-\Vgthree$ of Device~B for the same temperature values as in (a). Gate current $\Igtwo$ as a function of gate voltage difference $\Vgtwo-\Vgthree$ measured at $T=20~\mathrm{mK}$. (e) Critical current $ \Ic $ of Device~C as a function of gate voltage $ \Vgone $ for the same temperature values as in (a). (f) Gate current $\Igone$ of Device~C as a function of $\Vgone$ at $T=20~\mathrm{mK}$.}
 \label{fig:S2}
\end{figure}

In the Main Text we show a summary of the measurements obtained with Devices~B and~C using parametric plots of $\Ic $ as a function of $\Igone$ and $\Igtwo$. Figure~\ref{fig:S2} shows the datasets from which these parametric plots are obtained. Measurements were performed at temperatures ranging from $20~\mathrm{mK}$ (blue) to $3~\mathrm{K}$ (red). Gate currents are reported only for $20~\mathrm{mK}$.

\setcounter{myc}{2}

\section{Supplementing Information~3: Fit of the Switching Probability Distribution}
\setcounter{myc}{3}
\begin{figure}
 \includegraphics[width=\columnwidth]{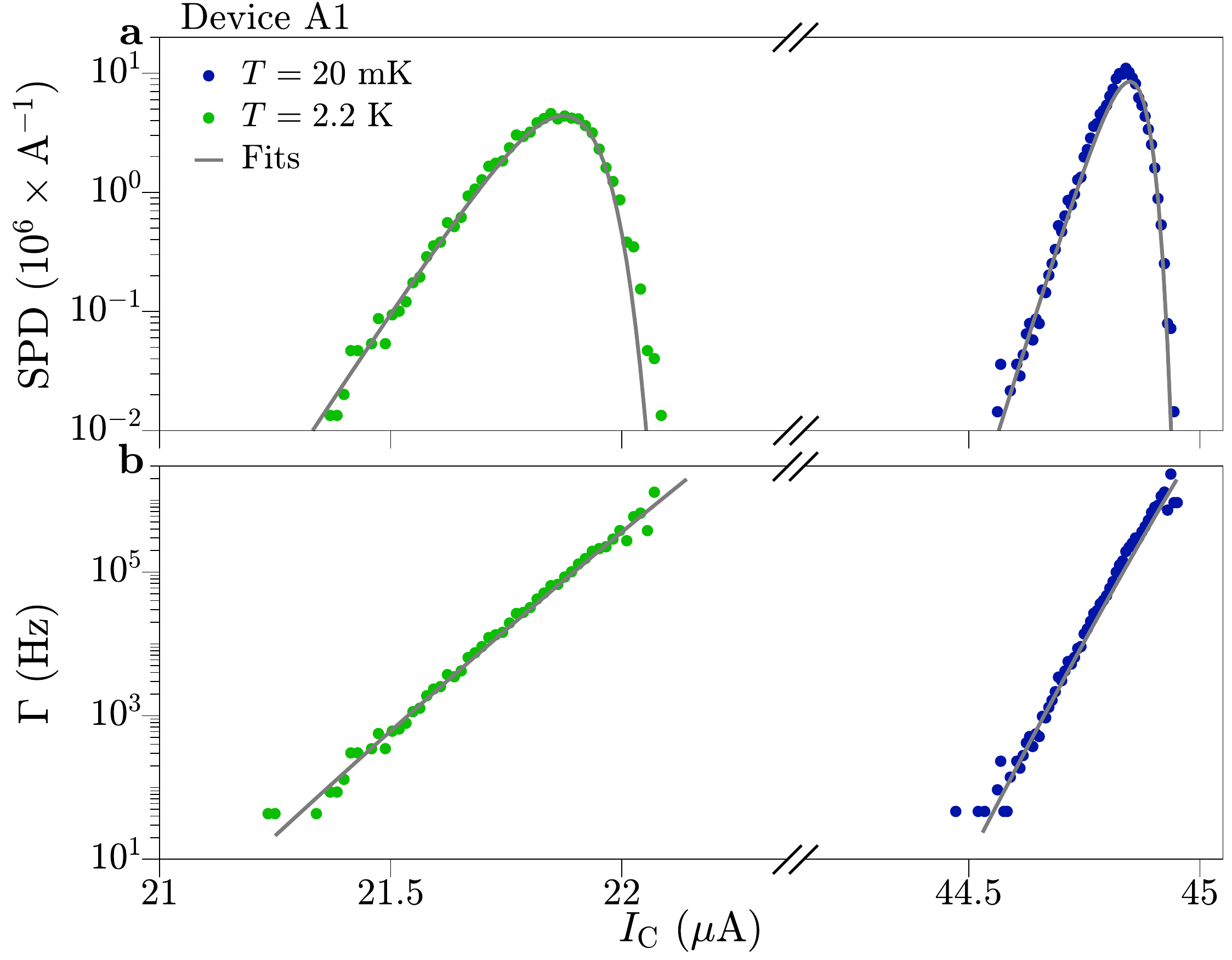}
 \caption{(a) Switching probability distribution in Device~A1, measured at temperatures $T$ of $20~\mathrm{mK}$ (blue circles) and $2.2~\mathrm{K}$ (green circles) together with fits of a theory of phase escape via macroscopic quantum tunneling and thermal activation (solid gray lines). (b) Phase particle escape rates of the same data as in (a), calculated with Eq.~\ref{eq:S3} (markers), together with calculations of the escape rate using Eq.~\ref{eq:S1}, using the fit parameters obtained from (a).}
 \label{fig:S3}
\end{figure}

Figure~4 of the Main Text shows the switching probability distribution (SPD) measured in Devices~A1 and C. Such measurements were performed by ramping the source drain current $\Isd$ and recording, for each sweep, the $\Isd$ value where switching from superconducting to resistive state occurred. Data was acquired for $2\times10^4$ switching events and $\Isd$ was ramped at a rate $v=6.4~\mathrm{mAs^{-1}}$. In Fig.~\ref{fig:S3}(a) we plot again the SPDs measured in Device~A1 at zero gate voltages and at a temperature of $20~\mathrm{mK}$ (blue circles) and $2.2~\mathrm{K}$ (green circles). In Fig.~\ref{fig:S3}(b) we plot the switching rates (markers), obtained from the data in Fig.~\ref{fig:S3}(a) by KFD transform~\cite{Bezryadin2012}: 
\begin{equation}
\Gamma(\Isd)=v P(\Isd)\left( 1-\int_{0}^{\Isd} P(I)dI\right)^{-1},
\label{eq:S3}
\end{equation}
where $P$ is the measured switching probability.

We fit to the SPD for each temperature, as shown in Fig.~\ref{fig:S3}(a), to a model for the switching rate of a superconducting nanowire~\cite{Murphy2013} that follows the relation:

\begin{equation}
\Gamma(\Isd,T)=\Omega(\Isd,T)\left[e^{-U(\Isd,T)/T} + e^{-U(\Isd,T)/\Tq}\right],
\label{eq:S1}
\end{equation}

where the attempt frequency is $\Omega(\Isd,T) = \Omega_{0} \left(1 - T^{2}/\Tc^{2}\right)^{3/4} \left(1 - \Isd/\Ic(T)\right)^{\nu}$ and the potential barrier height is $U(\Isd,T) = \frac{\kappa \hbar \Ic(T)}{e} \left(1-\Isd/\Ic(T)\right)^{\eta}$. $\Ic$ is considered to follow Bardeen's formula: $\Ic(T) = \Ico \left(1 - T^{2}/\Tc^{2}\right)^{3/2}$. The temperature dependence of the attempt frequency follows from $\Omega(\Isd,T) \propto \Ic^{1/2}$. For a nanowire forming a phase slip junction, we have $\kappa=\sqrt{6}/2$, $\eta=5/4$ and $\nu=5/8$. This model accounts for switching due to macroscopic quantum tunneling (MQT) and thermally activated phase escape mechanisms, which dominate at low and high temperatures respectively. We convert the modeled switching rate $\Gamma$ to a SPD using the inverse KFP transform~\cite{Bezryadin2012}:

\begin{equation}
P(\Isd) = \frac{\Gamma}{v} \mathrm{exp}\left(-1/v \int_{0}^{\Isd} \Gamma(I)dI \right).
\label{eq:S2}
\end{equation} 

We fit the logarithm of the probability distribution to optimally account for the shape of the SPD tails. We set $\Tc=3.7~\mathrm{K}$, as measured in Ref.~\cite{Ritter2021}, and fit with $\Omega_{0}$, $I_{c0}$ and $\Tq$ as free parameters. From the fit, we obtain $\Omega_{0} = 67.5\times10^{12}~\mathrm{rad s^{-1}}$, $I_{c0} = 45.7~\mathrm{\mu A}$, $\Tq=0.77~\mathrm{K}$. The finding of $\Tq\gg 20~\mathrm{mK}$ confirms MQT is the dominant phase escape mechanism at low temperatures. With these parameters, one should be able to calculate the SPD at any given temperature. However, we find that the curve at $T=2.2~\mathrm{K}$ is satisfactorily reproduced only by setting the parameter $\kappa$ to $71\%$ of its theoretical value, similar to previous observations~\cite{Murphy2013}. The value $\kappa=\sqrt{6}/2$ was derived under the assumption of a nanowire width much larger than the superconducting coherence length~\cite{Tinkham2002}. This assumption might not be completely justified in the current experiment.

\section{Supplementing Information 4: Reference Devices After Additional Fabrication}
\setcounter{myc}{4}
\begin{figure}
 \includegraphics[width=\columnwidth]{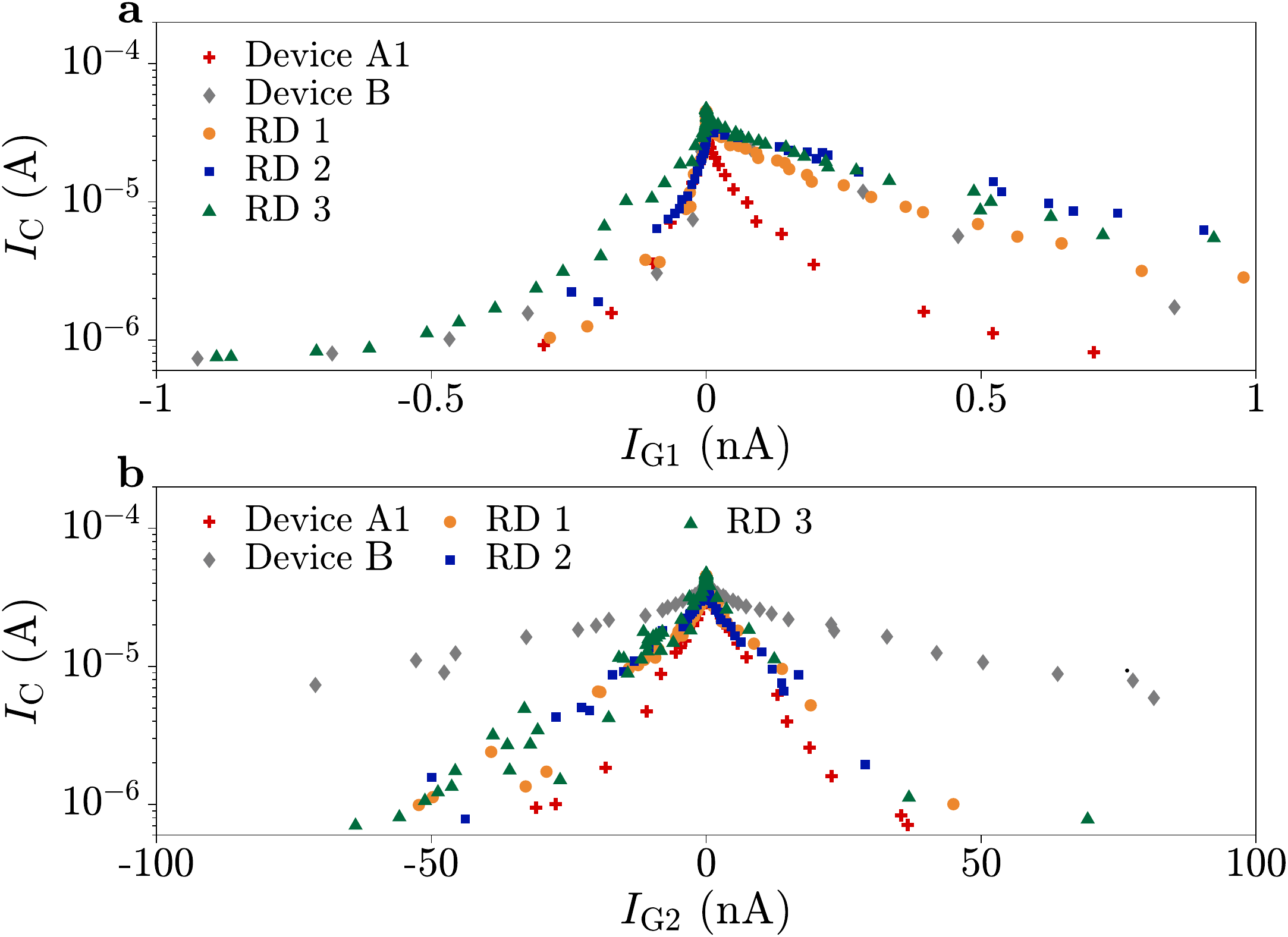}
 \caption{(a) Critical current $\Ic$ as a function of gate current $\Igone$ for several devices. All devices except Device~A1 underwent additional fabrication. (b) Parametric plot of critical current $\Ic$ as a function of gate current $\Igtwo$ for several devices. Device~B is the only one with a trench between gates and nanowire.}
 \label{fig:S4}
\end{figure}

In the Main Text we noted that devices which underwent additional fabrication steps, showed changes in some of their properties. Here we discuss this in more detail. In order to etch the trench into the Si substrate (see Fig.~2(a) of the Main Text) the entire sample was covered by a hard mask comprising a 2~nm thick $\mathrm{Si_3N_4}$ layer, grown by plasma enhanced atomic layer deposition, and a 210~nm thick $\mathrm{SiO_2}$ layer grown by plasma enhanced chemical vapor deposition. Both depositions were performed at a temperature of $300~^{\circ}\mathrm{C}$. After definition of the trench by electron beam lithography, reactive ion etching and inductively coupled plasma etching, the hard mask was removed by immersion in buffered HF. While the trench was etched only in Device~B, additional reference devices (RDs) on this chip underwent the same deposition and etching of the hard mask. We refer to these RDs as RD~1, RD~2 and RD~3, respectively. Reference devices had a similar geometry to Device~A1, with $d=1~\mathrm{\mu m}$, $800~\mathrm{nm}$ and $400~\mathrm{nm}$, respectively.

Figure~\ref{fig:S4}(a) shows a parametric plot of $\Ic$ as a function of $\Igone$ for all devices that underwent deposition and etching of the $\mathrm{Si_3N_4/SiO_2}$ hard mask, plus Device~A1. The trench in the Si substrate was etched only for Device~B. All devices that underwent further processing showed a characteristic asymmetric behavior, with $\Ic$ decreasing faster for $\Igone<0$ than for $\Igone>0$. We also found that RDs exhibited a reduced suppression efficiency for $\Igone>0$. Figure~\ref{fig:S4}(b) shows a parametric plot of $\Ic$ as a function of $\Igtwo$. We notice that Device~A1, which did not undergo additional fabrication, showed the fastest suppression of $\Ic$. Reference Devices~1,~2 and~3 have quantitatively similar behavior, despite the fact that $d$ varies from $400~\mathrm{nm}$ to $1~\mathrm{\mu m}$. This is presumably due to natural sample-to-sample variations following the additional fabrication.  It makes extraction of a dependence on $d$ difficult with just these three RDs. However, Device~B clearly stands out from the rest, indicating that the presence of the etched trench in the substrate causes a significant suppression of the long distance coupling between current $\Igtwo$ and nanowire. This clearly substantiates our explanation based on phonons.

\end{document}